\documentclass[prl,twocolumn,showpacs,superscriptaddress,amsmath,amssymb]{revtex4}
\usepackage[mathscr]{euscript}
\usepackage{cancel}
\usepackage{maybemath}
\usepackage{xcolor}
\usepackage{graphicx}

\begin{document}
THIS PAPER SHOULD HAVE BEEN WITHDRAWN
\title{Is the observability of sterile neutrino masses consistent with $\nu-$oscillations?}

\author{R. Ehrlich}

\affiliation{Department of Physics, George Mason University, Fairfax, Virginia 22030, USA}

\date{\today}

\begin{abstract}
It is shown that a 1980 proposal to search for heavy sterile neutrinos by observing the energy of the associated lepton in weak decays rests on an questionable assumption, and that the possibility of such a detection would be inconsistent with the observability of neutrino oscillations.
\end{abstract}
\maketitle
%
%
\section{Introduction}
In 1980 R. E. Shrock first proposed the idea of looking for heavy sterile neutrinos that might be a component of neutrino flavor states based on weak decay measurements including three body final states, e.g., beta decay as well as two body final states, e.g., $\pi \rightarrow \mu+\nu_\mu.$~\citep{Sh1980} 
Since that 1980 proposal many researchers have done experiments using Shrock's method,~\citep{Sc1983, Va1987, De1990, Si1985, Oh1993,Me2015} with some of them even reporting positive results, later proven incorrect. His method has also been accepted by many theorists and discussed in a number of well-regarded books.~\citep{Bo1987}  The purpose of the present paper is to note that the Shrock method, despite being an established understanding of the community, will not work based on simple theoretical grounds.  

Shrock argued that the admixture of mass eigenstates making up a flavor state no longer make up a single effective mass due to the large mass splitting when one of the mass states is sufficiently heavy.  As applied to beta decay for example it is claimed that the observed spectrum will consist of a sum of two spectra -- one for the heavy sterile neutrino $\nu_S$ and one for the other three ``light" mass states $\nu_L\equiv \nu_1, \nu_2,$ or $\nu_3,$ which may be treated as having a single effective mass in view of their small mass splittings.  The combined spectrum would then be found to have a kink in it which could be observed if the mixing angle between the sterile neutrino and the other mass states is not too small and if the statistics are sufficient.

\section{Decays into two-body final states}
The basis of Shrock's idea (and this authors criticism of it) can be most starkly illustrated for the case decays having two body final states such as $\pi \rightarrow \mu + \nu_\mu.$  Let us suppose Shrock were correct about the mass $m_S$ of any sterile neutrino component of $\nu_\mu$ being detectable if it were suffciently large based on decay kinematics.  Note that in such two-body decays it would not be necessary to observe the shape of a spectrum, since for a pion decay at rest the mass of the emitted neutrino is directly computable for \emph{individual} decays given the measured muon total energy $E_\mu=K_\mu+m_\mu c^2,$ and the known $m_\mu$ and $m_\pi.$  The relation based on simple kinematics is
\begin{equation}
m^2_\nu=m^2_\pi+m^2_\mu-2m_\pi E_\mu/c^2
\end{equation}

Since the muon is nonrelativistic, with $K_\mu=p_\mu^2/2m_\mu,$ and since $|p_\mu|= |p_\nu|$ then Eq. 1 can also be expressed as
\begin{equation}
m^2_\nu=A - Bp_\nu^2
\end{equation}

where A and B are constants dependent on $m_\mu$ and $m_\pi,$ and $p_\nu=p_\mu$ is known from the measured $K_\mu.$  

Let us imagine that a heavy sterile state $|\nu_S>$ existed having a mixing angle $\alpha=10^{-6}$ with the light states $|\nu_L>.$  According to Shrock, in an experiment observing $\pi \rightarrow \mu + \nu_\mu,$ one would find that in nearly all decays the emitted neutrino mass was close to zero, but for one decay out of $10^{12}$ Eq. 2 would yield the value for $m_\nu=m_S.$  Of course, such a result is what one would expect if the rate for the process were an incoherent sum of rates for $\nu_L$ and $\nu_S$, which would apply if those two final states are experimentally distinguishable for individual decays.   

The fallacy here is that the $\nu_L$ and $\nu_S$ final states are experimentally distinguishable \emph{only} if it were true that for some decays $\nu_\mu$ is in the final state $\nu_L$ while for others it it is in the $\nu_S,$ state, but it would not be true if $\nu_\mu$ is always born as a coherent mixture of both of these mass states.  In such a case, the rates for $\pi \rightarrow \mu + \nu_L$ and $\pi \rightarrow \mu + \nu_S$ must be added coherently, thereby yielding monochromatic muons whose energy would depend on the $\nu_\mu$ effective mass.  \emph{If one were to argue that the basic hypotheses of quantum mechanics require that individual emitted $\nu_\mu$ must be in a mass eigenstate (either $\nu_L$ or $\nu_S$) such a result would make it impossible for the $\nu_\mu$ to undergo subsequent oscillations, since that requires at least two nondegenerate mass states!}  Moreover, since the $\nu_L$ state really represents either $\nu_1,\nu_2$ or $\nu_3,$ the requirement that $\nu_\mu$ be emitted in a single mass eigenstate would make oscillations of the emitted $\nu_\mu$ unobservable between any pair of mass states, not just $\nu_S$ and $\nu_L$.  

Let us consider a final possibility that the unobservability of $\nu_\mu$-oscillations might be restricted to cases where its mass is first measured in say $\pi-$decay.  Although it must be acknowledged that issues of quantum entanglement need to be seriously considered,~\citep{Co2009} the preceding possibility may be untenable because one would imagine that the sum of rates over mass states must be either coherent or incoherent; the choice cannot be affected by a decision made \emph {subsequent} to the decay to measure $K_\mu$.  If the sum is coherent $\nu$-oscillations are observable, while if it is incoherent a sufficiently heavy $m_S$ is observable, while in no case are both observable.  

Writing in 1980, Shrock may have been unconcerned that his proposal meant that $\nu$-oscillations would be unobservable, but it certainly should be a matter of concern to researchers using the Shrock method to search for a heavy $m_S$ after $\nu$-oscillations were observed in 1998.   If the unobservability of neutrino oscillations were not enough to cast doubt on Shrock's claim, there are three further arguments against his notion that \emph{one must use an incoherent sum of rates over mass states if the weak gauge group (flavor) eigenstates are a combination of massless and massive (nondegenerate) neutrinos.}  (The preceding words in italics states Shrock's claim in one sentence, although it is not an exact quote from ref.~\citep{Sh1980}.)  

\begin{itemize}
\item Shrock makes the above claim of incoherence of the sum of rates without any proof or reference to other sources.
\item There is no reason to believe that Shrock's 1980 formulation of a combination of ``massive and massless neutrinos," applies to the current real world situation of a combination of ``light" and ``heavy" $\nu$ mass eigenstates (heavy = sterile), since no neutrinos are considered to be massless any longer.
\item Even if the combination of light and heavy neutrinos somehow did justify the use of an incoherent sum, this would imply that for some minimum mass difference $m_S-m_L$ (or ratio $m_S/m_L$) the sum was coherent.  There is then the awkward issue of explaining how such a value possibly could be defined.
\end{itemize}

\subsection{The double well analogy}
The relation between mass and flavor states is very similar to the situation of an electron in a symmetric double well potential.  Suppose the electron is known to be localized in one of the wells.  In such a case the electron's wavefunction can be expressed in terms of equal contibutions of the symmetric and antisymmetric energy eigenfunctions:  

\begin{equation}
|\psi>=(|\psi_S>+|\psi_A>)/\sqrt{2}
\end{equation}

whose energy eigenvalues are respectively $E_S$ and $E_A.$  We can easily find the average energy of this mixed state from $E=<\psi|H|\psi>.$  Since the states $|\psi_S>$ and $|\psi_A>$ are orthonogonal, we obviously have the result

\begin{equation}
E=(E_S+E_A)/2
\end{equation}

Now suppose there were a molecule that had such a double well potential and the molecule after being emitted in a chemical reaction had one of its electrons in a \emph{position} eigenstate, i.e., it resided in one well or the other.  In such a case, if it were possible to measure the electron's energy based on the reaction kinematics that created the molecule we would find the value of its energy would be given by Eq.~4.  We would certainly not find $E_S$ and $E_A$ each half the time, again assuming the electron in the emitted molecule were known to be in a position eigenstate.  Of course, if we measured the electron's energy \emph{after} the molecule had been emitted we would indeed find either $E_S$ and $E_A$ each half the time.  In this analogy we may think of the position eigenstate as being akin to a $\nu$ flavor state, and the S and A energy eigenstates as being a pair of mass eigenstates.  

It is interesting that such double well molecules actually exist with ammonia ($NH_3$) being one example.  In $NH_3$ the three hydrogens form an equilateral triangle and the nitrogen atom, playing the role of the electron in the double well example, lies at the apex of a tetrahedron above the plane of the 3 H's.  The nitrogen atom can also sometimes be found in the other well (located at the ``anti-apex"), an equal distance below the plane.  In fact the nitrogen oscillates between those two positions with a frequency that depends on the barrier between the wells.

\subsection{Finding the $\nu_\mu$ effective mass}
As before we express the $|\nu_\mu>$ flavor state as a mixture of the light $(\nu_L)$ and sterile $(\nu_S)$ states.

\begin{equation}
|\nu_\mu> = \sin\alpha|\nu_S>+\cos\alpha |\nu_L> 
\end{equation}

Here $|\nu_S>$ and $|\nu_L>$ will be orthogonal eigenfunctions of the momentum operator $\mathcal{P}$ with eigenvalues $p_S$ and $p_L.$   The orthogonality of $|\nu_S>$ and $|\nu_L>$ holds because the following pair of equations must both be true:
\begin{equation}
<\nu_S|\mathcal{P}|\nu_L> = <\nu_S\mathcal{P}|\nu_L>=p_S<\nu_S|\nu_L>
\end{equation}
\begin{equation}
<\nu_S|\mathcal{P}|\nu_L> = <\nu_S|\mathcal{P}\nu_L>=p_L<\nu_S|\nu_L>
\end{equation}
which requires that $<\nu_S|\nu_L>=0$ since $p_L$ and $p_S$ are unequal.  The experimentally measured value of the $\nu_\mu$ momentum $p_\nu$ for any decay event can be expressed in terms of the expectation value of $\mathcal{P},$ which based on the orthogonality of $|\nu_S>$ and $|\nu_L>$ is
\begin{equation}
p_\nu=<\nu_\mu|\mathcal{P}|\nu_\mu> = p_S\sin^2\alpha  +p_L \cos^2\alpha
\end{equation}

and where based on Eq. 2, we have $p_S=\sqrt{(A-m_S^2)/B},$ and $p_L=\sqrt{(A-m_L^2)/B}.$ When these two relations are substituted in Eq. 8 the result is an implicit expression for the $\nu_\mu$ effective mass $m_\nu$ for the decay in terms of $m_L$ and $m_S.$

\begin{equation}
\sqrt{\frac{A-m_\nu^2}{B}} 
=sin^2\alpha \sqrt{\frac{A-m_S^2}{B}} + \cos^2\alpha \sqrt{\frac{A-m_L^2}{B}}
\end{equation}

whose solution for $m_\nu^2$ yields the expected result:
\begin{equation}
m_\nu^2=m_S^2 sin^2\alpha  + m_L^2\cos^2\alpha 
\end{equation}

Thus, a measurement of the $\nu_\mu$ effective mass $m_\nu$ obtained from a direct measurement of $K_\mu$ for each individual pion decay will take on the above value that is intermediate between $m_L$ and $m_S.$  It most certainly will not be $m_L$ for some decays and $m_S$ for others.

\section{Search for a heavy $\nu_S$ in $\beta$-decay}
Currently, the KATRIN experiment~\citep{Me2015} hopes to be able to measure the electron neutrino effective mass to an order of magnitude greater sensitivity than the present upper limit of 2 eV,~\citep{Ol2014} based on the spectral shape near the endpoint.  The $\nu_e$ effective mass in beta decay can be expressed in terms of the mass eigenstates comprising it.~\citep{Ol2014}
\begin{equation}
m_{eff}^2=\Sigma |U_{ei}|^2m_i^2
\end{equation}
 A secondary objective of the experiment is to look for evidence of a keV mass sterile neutrino $\nu_S$ contributing to $\nu_e.$  It is claimed that even with an active sterile mixing as small as $sin^2\theta=10^{-6},$ one should be able to detect a spectral kink whose position would reveal the mass of a sterile state.~\citep{Me2015}

In order to see why the spectral shape cannot reveal the presence of a heavy sterile neutrino let us start with Fermi's Golden Rule that describes the $\beta-$decay rate when the emitted electron has a kinetic energy $E.$ 
\begin{equation}
\frac{d\Gamma}{dE}_{coh}=\frac{2\pi}{\hbar}|M|^2\rho 
\end{equation}
where $\rho$ is a three-body phase space factor that depends on the $\nu_e$ effective mass (Eq. 11).  One could only justify separate $\rho_i$ (associated with the individual $\nu_i$) in Eq. 12 if the $\nu_e$ were born as one of the $\nu_i$ mass states comprising it, which as already discussed in the previous section is not the case.  
The $M$ in Eq. 12 is the matrix element for the decay, $M = <W^-|H|e^-\bar{\nu}_e>.$   Given that $|\bar{\nu}_e>=\Sigma U_{ei}|\bar{\nu}_i>,$ M can be written as
\begin{equation}
M =  <W^-|H|e^-\Sigma U_{ei}|\bar{\nu}_i>=\Sigma U_{ei}M_i
\end{equation}
with $M_i = <W^-|H|e^-\bar{\nu}_i>.$  Substitution of Eq. 13 into Eq. 12 then yields
\begin{equation}
\frac{d\Gamma}{dE}_{coh}= \frac{2\pi}{\hbar}|\Sigma U_{ei} M_i|^2\rho
\end{equation}

In Eq.~14 we have expressed $M$ as a linear sum of matrix elements $M_i$ with weights $U_{ei}.$  Since the terms being summed are added before squaring, the sum is a coherent one, and the decay rate $\frac{d\Gamma}{dE}$ therefore involves cross terms of the form $U^*_{ei}M^*_iU_{ej}M_j.$   

The basis of using a coherent sum here is much the same as in the previously discussed two-body weak decay.  Let us now suppose that the sum in Eq.~14 had been written incorrectly using an incoherent sum over the matrix elements:
\begin{equation}
\frac{d\Gamma}{dE}_{inc}= \frac{2\pi}{\hbar}\Sigma |U_{ei}|^2 |M_i|^2\rho_i
\end{equation}
It is clear from the form of Eq.~ 15 that it describes a spectrum consisting of a sum of separate spectra associated with each mass state with weights $|U_{ei}|^2\rho_i.$   This situation in fact is what Shrock and the authors of refs. 2-8 claim applies to $\beta-$ decay.  However, this would mean that individual neutrinos are emitted in specific mass states contributing to $\nu_e$, making it impossible for them to undergo oscillations.

\section{Possible ways to find $m_S$}
There are of course ways that an experiment might detect a sterile neutrino.  One possibility would be based on neutrino oscillations, as is being done in the Daya Bay~\citep{Dw2013} and DUNE experiments~\citep{Be2015} that are looking for $m_S^2$ on the order of $1 eV^2.$ Alternatively one might seek evidence for much heavier nearly degenerate sterile neutrinos based on the mixing and oscillations between them that might be seen as quantum beats in the distribution of final states.~\citep{Bo2014}  Evidence for both light and heavy sterile neutrinos has also been suggested  as possible dark matter candidates based on various models.~\citep{Ch2014, Ch2016}  

Moreover, if we were fortunate enough to have another galactic supernova at some distance $cT,$ and if $m_\nu \ge 1eV$ one should be able to deduce the masses of $\emph{individual}$ arriving neutrinos based on their energies $E_i$ and their travel time delays relative to light $\Delta t_i = T-t_i,$ based on relativistic kinematics:
\begin{equation}
m^2_i=\frac{2E_i^2\Delta t_i}{T}
\end{equation}

Note, that in such a case the neutrinos emitted from the core collapse are in distinct flavor states, whose mass states comprising them lose coherence after travelling a distance greater than the coherence length, as discussed by Giunti and Kim, and they can be observed as mass states on reaching Earth.~\citep{Gi1998}  

The author is compelled to conclude this paper on a strange note in order to avoid confusion on the part of readers who may have read another recent paper by him in which he makes use of the conventional formula (Eq. 15) to describe the $\beta$ spectrum.~\citep{Eh2016}  Despite the arguments presented here, it is possible that the conventional (Shrock) treatment is in fact the correct one, given the quantum entanglement argument noted earlier.  In fact, given the uniqueness of the prediction made in ref.~\citep{Eh2016} for the results of the KATRIN experiment, the author sincerely hopes that the argument made in the present paper is proven to be incorrect!
\\
\section{acknowledgment}
The author wishes to thank Alan Chodos for helpful comments.


\end{document}